\documentclass[aps,twocolumn,groupedaddress,showpacs]{revtex4}
\usepackage{amsmath}
\usepackage{amssymb}
\usepackage{amsfonts}
\usepackage[dvips]{graphicx}
\usepackage[]{caption}
\usepackage[]{epsfig}
\begin{document}

\title{Reduction of the hydrophobic attraction between charged solutes
in water}

\author{J. Dzubiella}
\email[e-mail address: ] {jd319@cam.ac.uk}
\affiliation{University Chemical Laboratory,
Lensfield Road,
Cambridge CB2 1EW,
United Kingdom}
\author{J.-P. Hansen}
\affiliation{University Chemical Laboratory,
Lensfield Road,
Cambridge CB2 1EW,
United Kingdom}
\date{\today}

\begin{abstract}
We examine the effective force between two nanometer scale solutes in
water by Molecular Dynamics simulations. Macroscopic considerations
predict a strong reduction of the hydrophobic attraction between
solutes when the latter are charged. This is confirmed by the
simulations which point to a surprising constancy of the effective
force between oppositely charged solutes at contact, while like
charged solutes lead to significantly different behavior between
positive and negative pairs. The latter exhibit the phenomenon
of ``like-charge attraction" previously observed in some colloidal
dispersions. 
\end{abstract}

\pacs{82.70.Uv,87.16.Ac,61.20Ja,68.08.Bc}

\maketitle
When apolar solutes are dispersed in water, they exert effective,
solvent induced forces on each other, generally referred to as
hydrophobic attraction. For large solutes, say globular proteins, the
mechanism for hydrophobic interactions can be traced back to solvent
depletion (or ``drying") from the volume bounded by opposite surfaces
of the solutes \cite{wallquist:jpc,lum:jpc}. This mechanism, which is
reminiscent of polymer-induced depletion attraction between colloidal
particles on larger (mesoscopic) scales \cite{likos:physrep}, is most
pronounced when the solvent is near liquid-vapour coexistence
\cite{lum:jpc}. However, many biomolecular solutes are charged, and in
this work we present evidence, based on extensive Molecular Dynamics
(MD) simulations, that the electric field due to the charges carried
by the solutes leads to a considerable reduction of the hydrophobic
attraction. 

Such a reduction may be qualitatively understood from macroscopic
considerations. It is a well-known fact that a polar liquid, like
water, will rise inside a vertical condenser to minimize the overall
electrostatic energy (the rise being limited by gravity on a
macroscopic scale) \cite{landau}. Transposing this situation to the
nanometric scale, consider two parallel plate-like solutes of area
$A$, separated by a distance $D$, inside a polar solvent of relative
dielectric permittivity $\epsilon$, carrying opposite surface charges
$\pm\sigma$. Neglecting edge effects, the electric field between the
plates is $E_{0}/\epsilon$ with $E_{0}=\sigma/2\epsilon_{0}$. We require the
difference in the grand potential between the situations where the liquid
solvent ($l$) or its vapour ($g$) fill the volume $AD$ between the
two plates:
\begin{eqnarray}
\Omega_{\alpha}=-P_{\alpha} AD + 2\gamma_{w\alpha}A +
\frac{1}{2}\epsilon_{0}\frac{E^{2}_{0}}{\epsilon_{\alpha}}AD;\;\;\;\;\; \alpha=l,g
\end{eqnarray}
where $P_{\alpha}$ is the pressure of phase $\alpha$ and
$\gamma_{w\alpha}$ the surface tension between phase $\alpha$ and the
plate (``wall"). Consider a state close to phase coexistence at
temperature $T$, and let $\delta\mu=\mu-\mu_{\rm sat}$ be the positive
deviation of the chemical potential from its saturation value. Expanding
the $P_{\alpha}$ to linear order in $\delta\mu$ around their common
value at saturation, one easily arrives at the following expression for
the difference in grand potentials per unit area:  
\begin{eqnarray}
\frac{\Omega_{l}-\Omega_{g}}{A}&=&(\rho_{ g}-\rho_{
  l})\delta\mu D+2(\gamma_{ w l}-\gamma_{w
  g})\\&+&\frac{\epsilon_{0}}{2}E^{2}_{0}(\frac{1}{\epsilon_{
  l}}-\frac{1}{\epsilon_{g}})D \nonumber
\end{eqnarray}
At the ``drying" transition between the plates,
$\Omega_{l}-\Omega_{g}=0$ and
$\gamma_{wl}-\gamma_{wg}=\gamma_{lg}\equiv\gamma$. Since
$\rho_{g}\ll\rho_{l}$, and
$\epsilon_{l}\equiv\epsilon\gg\epsilon_{g}\simeq 1$, eq. (2) yields
the following expression for the critical distance $D_{\rm c}$ between the
plates at which drying occurs:
\begin{eqnarray} 
D_{\rm c}\simeq\frac{2\gamma}{\rho_{l}\delta\mu+\frac{\epsilon_{0}}{2}E^{2}_{0}}
\end{eqnarray}
For strong electric fields, ($E_{0}\lesssim 10^{10}$V/m corresponding
to surface charges $\sigma\lesssim e/{\rm nm}^{2}$), the electrostatic
term in the denominator is typically 10 times larger than the
$\delta\mu$ term in the vicinity of gas-liquid coexistence, and
leads to values of $D_{\rm c}$ of the order of a few \AA. This strong
reduction of $D_{\rm c}$ hints at a considerable weakening of the hydrophobic
interaction between two solutes when the latter are charged. Note that
within our macroscopic model this reduction is due to the overall
electric field, not to any Hydrogen-bonding of the solvent molecules
to hydrophilic ``patches" on the solute surface.

In order to confirm the qualitative prediction of the schematic model,
we have carried out extensive MD simulations of two spherical solutes
immersed in a bath of SPC/E water molecules \cite{berendsen:jpc}. The
solutes are spheres of radius $R$ which repel the solvent
molecules by a repulsive $\epsilon (r-R)^{-12}$, where $r$ is the distance from
the solute center to the oxygen atom of a water molecule;
 the energy scale $\epsilon$ is chosen such that the O atom
 experiences an energy $k_{\rm B}T$ at a distance of $r-R=$1\AA~ from 
the solute surface. The simulation cell is a cube of length up to
$L=40$\AA, containing up to 2000 water molecules, depending on the
solute size; solutes are placed at fixed positions on the body diagonal of the
simulation cell. The box dimensions are chosen such that the surface
to surface distance to the nearest image solute is at least 20\AA. 

The MD simulations are carried out with the DLPOLY2 package
\cite{dlpoly}, using the Verlet algorithm \cite{frenkelsmit}, with a
timestep of 2fs. The Berendsen barostat and thermostat
\cite{berendsen:jcp} were used to maintain the SPC/E water at a
pressure of 1 bar and a temperature $T=300$K. All electrostatic
interactions were calculated using particle-mesh Ewald summations
\cite{essmann:jcp}. 
\begin{figure}
  \begin{center}
\includegraphics[width=8.6cm]{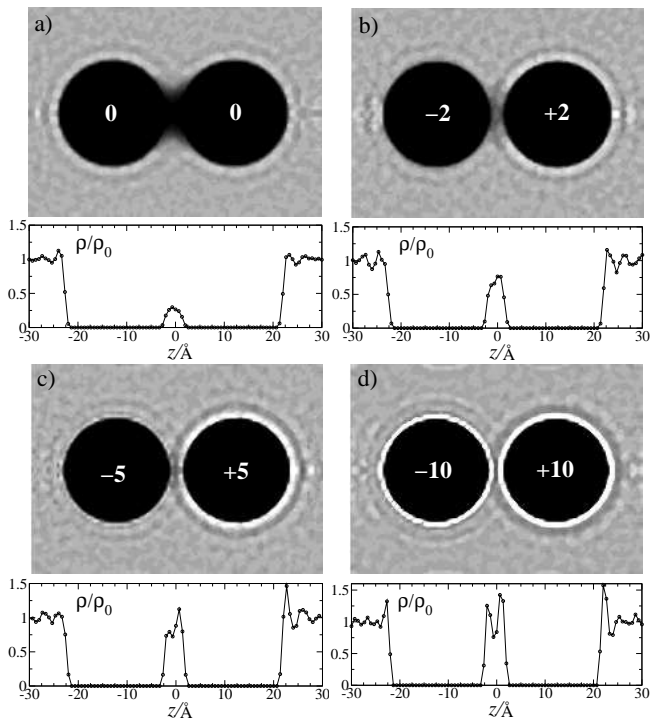}
    \caption{Density profiles of the water molecules around two
      neutral spherical solutes of radius $R=10$\AA~ (a) and two oppositely
      charged spherical solutes of radius $R=10$\AA~ carrying a
      charge $\pm qe$ with (b) $q=2$, (c) $q=5$, and (d)
      $q=10$. The surface-to-surface distance in all cases is
      $s=4$\AA. In the contour plots  dark regions show low density
      regions while high densities are plotted bright. The panels
      below the contour plots show the water density $\rho$ scaled
      with water bulk density $\rho_{0}$ in a cylinder of radius
      $R_{c}=5$\AA, coaxial with the center-to-center line of the
      solutes.}    
\label{fig1}
\end{center}
\end{figure}
We first consider the case of uncharged solutes. The water density
profile is illustrated in frame (a) of Fig. 1 for the case of solutes
of radius $R=10$\AA~ and a surface-to-surface distance along the $z$-axis
joining the centers $s=4$\AA. The upper part of the frame shows
a density contour plot coded by variable shades of grey. The lower part
shows density profiles along the center-to-center to axis $z$, averaged over a
coaxial-cylindrical volume of radius $5$\AA. The density profiles
show a considerable depletion of the solvent within a radial distance
of 5\AA~ from the center-to-center axis, reminiscent of the
observations of Wallquist 
and Berne for flatter solutes \cite{wallquist:jpc}.  As the
surface-to-surface distance $s$ is increased for fixed radius $R$, the
water molecules penetrate into the region between opposite solute
surfaces, as signalled by a rapid increase of the central peak (around
$z=0$) in the density profiles. When $s\approx6.5$\AA, the solvent
layers around an isolated solute are hardly disturbed by the presence
of the other solute.

\begin{figure}
  \begin{center}
\includegraphics[width=8cm,angle=0.,clip]{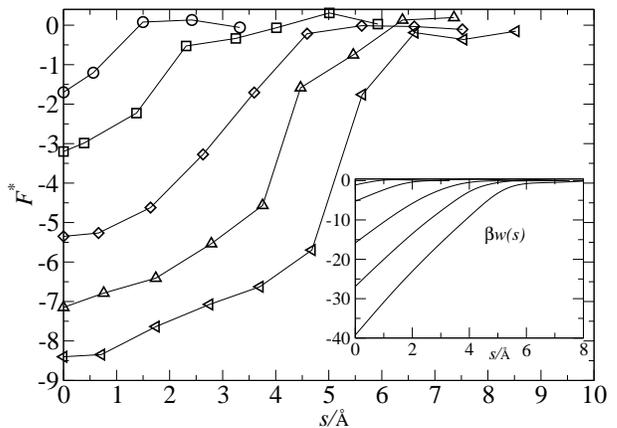}
    \caption{Simulation results (symbols) of the mean force,
      $F^{*}=\beta F$\AA, between
      two neutral ($q=0$) spheres in SPC/E water. Error bars 
      are omitted for clarity; the error is estimated to be $\Delta
      F^{*}=0.4$, comparable to the symbol size. The
      lines are guide to the eye. Results are plotted for solute
      radii  $R=3$\AA~ (circles), $R=5$\AA~ (squares), $R=8$\AA~
      (diamonds), $R=10$\AA~ (triangles pointing up), and $R=12$\AA~ (triangles
      pointing left). The inset shows the integrated force (potential of mean
      force) in obvious order.}  
\label{fig2} 
\end{center}
\end{figure}
The mean effective force acting on each of the solutes in the presence
of the second at a surface-to-surface distance $s$ is calculated by
averaging the total force due to all solvent molecules over the
configurations generated by MD runs extending over typically 1ns. This
average force obviously goes to zero at large distances $s$ and for
symmetry reasons, it is directed along the center-to-center
axis. Examples for several radii 3\AA~$\leq R \leq 12$\AA~ are shown in
Fig. 2. The largest radii are of the order of the size of small
globular proteins or of oil-in-water micelles. As expected from a
depletion mechanism, the force is attractive and its contact values
and range increase with $R$. The potentials of mean force $w(s)$ may
be calculated for each $R$ by integrating the force. The resulting
potentials are shown in the inset to Fig. 2. They closely resemble
results obtained for polymer-induced depletion potentials between
spherical colloids, albeit on different length and energy scales
\cite{wallquist:jpc,louis:jcp}. Note that the force at contact,
$F(0)$, scales roughly with $R$. This may be rationalized by a simple
consideration of the potential of mean force for plates ($R=\infty$)
at contact, $w(0)=-2\gamma$ (where $\gamma$ is now the plate-solvent
surface tension which differs little from the liquid-gas surface
tension \cite{huang:jpc}) and an application of the Derjaguin
approximation, valid for weakly curved substrates (i.e. large $R$)
\cite{louis:jcp}; this leads to the estimate $F(s=0)=-2\pi R\gamma$,
which indeed predicts linear scaling; the value $\gamma\simeq 0.05\,{\rm
J/m^{2}}$, extracted from the MD value of $F(0)$ for the largest $R$,
is reasonably close to the liquid-vapour surface tension of  
water under normal conditions ($\gamma=0.073 \,{\rm J/m^{2}}$). The
agreement must be considered quite satisfactory in view of the
roughness of the estimate. 

\begin{figure}
  \begin{center}
\includegraphics[width=8cm,angle=0.,clip]{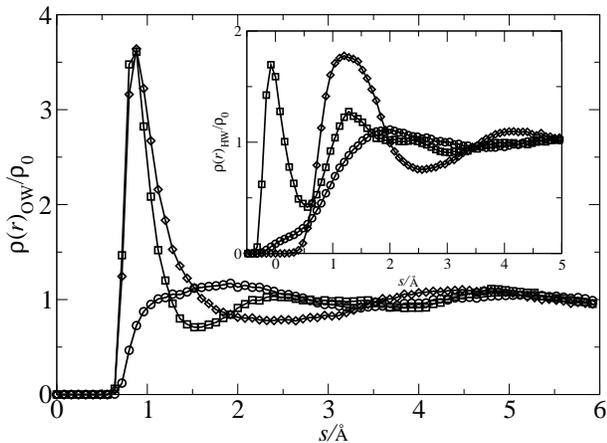}
    \caption{Density profiles of water oxygen and hydrogen atoms
      (inset) around one isolated solute with radius $R=10$\AA~ and central charge $q=0$
      (circles), $q=-10$ (squares), and $q=10$ (diamonds).}
\label{fig3}
\end{center}
\end{figure}

\begin{figure}
  \begin{center}
\includegraphics[width=8cm,angle=0.,clip]{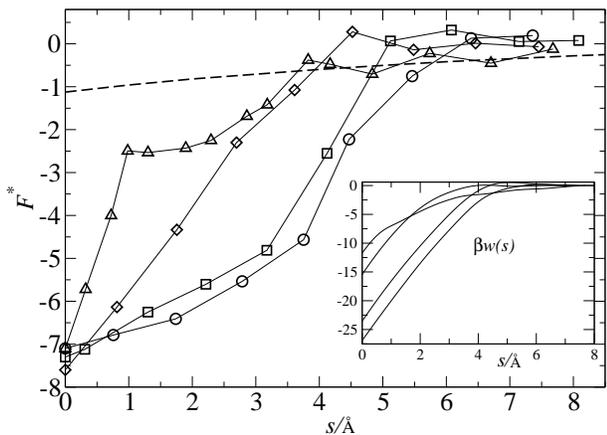}
    \caption{Same as Fig. 2 but now the mean force, $F^{*}=\beta F$\AA,
      is for oppositely charged spheres of
      $R=10$\AA~ and different central charges $\pm qe$ are shown:
      $q=0$ (circles), $q=2$ (squares), $q=5$ (diamonds), $q=10$
      (triangles up). The dashed line represents the electrostatic
      force between 2 periodically repeated solutes with opposite
      charges $q=\pm$ 10 in a continuous solvent with permittivity
      $\epsilon=80$.  The inset shows the resulting potentials of mean
      force; the contact values $w(s=0)$ increase with $q$.}
\label{fig4}
\end{center}
\end{figure}
We now turn to the charged solutes. Frames (b)-(d) in Fig. 1 show
water density profiles in the vicinity of two spheres carrying
opposite electric charges $\pm qe$ at their center (opposite charges
ensure overall charge neutrality without any need for counterions). As
$q$ increases from zero (frame (a)), water is seen to penetrate between
the two solutes, the central peak around $z=0$ in the density profiles
increases rapidly and its amplitude reaches roughly the bulk density
of water when $q=10$. Note that this central peak is asymmetrically
split, indicating the presence of two hydration layers which differ
somewhat depending on their association with the anionic or cationic
solute. This difference is also evident in the contact values of the
outside surfaces of the solutes, and is a consequence of
the different arrangements of the water dipoles around the solutes
induced by the local electric fields. The asymmetry of the profiles
can be rationalized by inspecting the density profiles of O and H
atoms around isolated solutes, plotted in Fig. 3. The hydration shell 
is more sharply defined  around the cationic than around the anionic
solute. The water dipoles tend to point radially away from the cation,
while the opposite configuration is more favourable around
anions. Note also that driving water into a narrowly confined region
under the action of a strong electric field (here between oppositely
charged solutes) is an effect reminiscent of that observed 
in recent MD simulations of ion permeation of hydrophobic nanopores
\cite{joe:channel}.  

The resulting mean forces between solutes are plotted for $q=0,2,5$
and 10, as functions of the surface-to-surface distance $s$ in
Fig. 4 together with corresponding potentials of mean force. The mean
force includes the direct Coulomb interaction between the two solutes
(with proper account for the periodic images), which is in fact an
order of magnitude larger that the total mean force. At large
distances hydrophobic interactions become negligible and the force
should tend to $-q^{2}e^{2}/(4\pi\epsilon_{0}\epsilon r^{2})$, where
$r=2R+s$ and $\epsilon$ is the dielectric permittivity of bulk SPC/E
water; the corresponding curve is also shown in Fig. 4.

\begin{figure}
  \begin{center}
\includegraphics[width=8cm,angle=0.,clip]{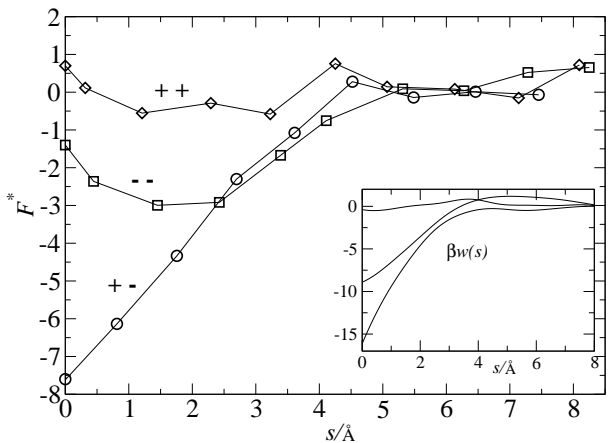}
    \caption{Same as Fig. 2 but now the mean force, $F^{*}=\beta
      F$\AA, is for different signs of
      the  charged spheres of radius $R=10$\AA~ and fixed central charges
      $qe=\pm 5e$ are shown: oppositely charged (+ --,circles), equally
      charged with $q=-5$ (-- --,squares), equally charged with $q=5$
      (++,diamonds).}  
\label{fig5}
\end{center}
\end{figure}
The most striking result illustrated in Fig. 4 is the near
independence of the force at contact, $s=0$, with respect to
solute charge. From the density profiles in Fig. 1 the hydrophobic
attraction is expected to be reduced but this reduction is almost
exactly compensated by the Coulomb attraction between solutes in the
presence of the solvent. As $q$ increases, the initial slope of the
effective force increases. The potential of mean force (shown in the
inset of Fig. 4) exhibits a contact value which increases with $q$,
indicating that the reduction of hydrophobic attractive free energy
and the reduction of dielectric screening due to restructuring of
water by the strong electric field of the solutes clearly outweigh the
increase in bare Coulomb attraction between the latter. Simulations
calculating the forces at and  near contact for $q=7$ and $q=15$, not 
shown in Fig. 4, confirm this trend. Note that the potential of
mean force curve for $q=10$ shows more long range attraction as
compared to the smaller $q$ data due to the increased electrostatic 
attractive interaction.The eye-catching kink in the force for $q=$10,
at a distance $s\approx 1$\AA~ is reproducible, and is probably
related  to the pronounced shell structure of water molecules around
highly charged solutes, illustrated in Fig. 3. While for neutral (and
weakly charged) solutes, the O and H density profiles show little
structure, they are sharply peaked at a distance  $s\approx 1$\AA~ of
the O atoms from the solute surface. This would lead to a complete
shared hydration layer, and consequently to a kink in the force versus
distance curve, between 2 flat solutes separated by $s=2$\AA~. This
critical separation is shifted to shorter distances due to 
the curvature of spherical solutes. 
 
In view of this delicate balance between various interactions, we have
also examined the case of equally charged solutes. In this case
monovalent counterions (Na$^{+}$ or Cl$^{-}$) were included to ensure overall
charge neutrality. The situation is summarized in Fig. 5 for solutes
of radius $R=10$\AA~ and charge $q=\pm5$. The water density profiles
(not shown here) are symmetric with respect to $z=0$ for equally
charged solutes, but differ substantially when going from a pair of
anions to a pair of cations. In the latter case water is much more
structured into well defined hydration shells, as shown in
Fig. 3. This difference is 
reflected in the effective forces and potentials shown in
Fig. 5. While the force and potential of mean force are practically
zero at all distances in the case of cationic solutes, anionic solutes
attract each other significantly, showing that residual hydrophobic attraction
overcomes the repulsion between like charges, which may be more
efficiently reduced by the local permittivity of water in the
immediate vicinity of the anionic solutes. 

To summarize, we have shown that electric charges carried by
nanometer-scale solutes have a profound influence on hydrophobic
interactions. Our MD simulations confirm that for neutral solutes
hydrophobic attraction is intimately linked to solvent depletion, and
the contact value of the hydrophobic force is directly related to the
surface tension of the pure solvent. Solvent depletion is suppressed by
the electric field due to any charge carried by the solute, but for
oppositely charged solutes the resulting loss of hydrophobic
attraction is compensated by their mutual Coulomb attraction, leading
to a nearly constant contact force. The effective force between
equally charged solutes depends on the sign of their charge. Anionic
solutes exhibit a striking ``like-charge attraction", a subject of
considerable debate in the field of mesoscopic charge-stabilized
colloids \cite{hansen:review}. Coulombic and hydrophobic effects are intimately
entangled, and their inter-relation is expected to play an important
role in the analysis of protein aggregation and related biomolecular
mechanisms. 

The authors are grateful to R. Evans and H. L{\"o}wen for useful clarifications. JD
acknowledges the financial support of EPSRC within the Portfolio Grant
RG37352.

\end{document}